# Giant oscillator strength in the band-edge light absorption of zincblende, chalcopyrite and kesterite solar cell materials

Masato Kato, Mitsutoshi Nishiwaki and Hiroyuki Fujiwara[*]

*Department of Electrical, Electronic and Computer Engineering, Gifu University, 1-1 Yanagido, Gifu 501-1193, Japan*

## Abstract

In semiconducting solar-cell absorbers, high absorption coefficient ($\alpha$) near the band-edge region is critical to maximize the photocurrent generation and collection. Nevertheless, despite the importance of the band-edge absorption characteristics, the quantitative analysis of the band-edge optical transitions has not been performed. In this study, we have implemented systematic density functional theory (DFT) calculation, focusing on the band-edge oscillator strength of seven practical solar cell absorbers (GaAs, InP, CdTe, $CuInSe_2$, $CuGaSe_2$, $Cu_2ZnSnSe_4$, and $Cu_2ZnSnS_4$) with zincblende, chalcopyrite and kesterite structures. We find that all these crystals exhibit the giant oscillator strength near the band gap region, revealing the fact that $\alpha$ in the band gap region is enhanced significantly by the anomalous high oscillator strength. In high-energy optical transitions, however, the oscillator strength reduces sharply and the absorption properties are determined primarily by the joint density-of-state contribution. Based on DFT results, we show that the giant oscillator strength in the band edge region originates from a unique tetrahedral-bonding structure, with a negligible effect of constituent atoms.

[*]fujiwara@gifu-u.ac.jp.

## I. INTRODUCTION

Intense light absorption within solar-cell light absorbers is essential to achieve high photo-induced current and the resulting high photovoltaic performance [1,2]. To realize high optical gains, almost all practical solar cells employ direct-transition semiconducting absorbers that exhibit high absorption coefficient ($\alpha$). Nevertheless, total light absorption in semiconductors is often limited by low $\alpha$ values observed in the band edge region ($\alpha \sim 10^4$ cm$^{-1}$) [3-5]. Thus, strong band-edge absorption is desirable for the enhanced carrier generation, allowing the effective reduction of a solar-cell absorber thickness.

Rather surprisingly, despite the importance of the band-edge light absorption in photovoltaic devices, the efforts for the theoretical interpretation of absolute $\alpha$ in the band gap ($E_g$) region are scarce. In general, the light absorption in materials is discussed based on the joint density-of-state distribution [6,7]. However, optical transitions in semiconductors can be influenced strongly by oscillator strength ($f$), which represents the magnitude of the interaction between conduction- and valence-band wavefunctions upon polarized light excitation [8]. The absolute $f$ values in the $E_g$ region, which can further be related to band-edge $\alpha$ values, could be treated as critical values concerning light absorption in semiconductor materials. At this stage, only limited studies have been made to determine the $f$ characteristics of materials [9].

Quite fortunately, exact $f$ values can be evaluated directly by applying density functional theory (DFT) [9]. In fact, our earlier studies showed that the $\alpha$ spectra calculated by DFT reproduce the experimental $\alpha$ spectra of various photovoltaic materials almost perfectly [9,10], confirming the validity of the overall DFT calculations. These works have implied the possibility that quite detailed analyses of band-edge optical transitions can be made based on DFT. Such an approach is expected to open a new path for finding a superior high-$f$ solar cell material within the framework of DFT.

So far, zincblende, chalcopyrite and kesterite-based compound semiconductors have been studied extensively as solar cell materials. Specifically, GaAs, InP and CdTe have been applied as high-efficiency zincblende materials [11], whereas Cu(In,Ga)Se$_2$ (CIGSe) has been applied as a chalcopyrite semiconductor [12,13]. More recently, Cu$_2$ZnSnSe$_4$ (CZTSe) and Cu$_2$ZnSnS$_4$ (CZTS) kesterite crystals have been investigated intensively as emerging materials [14,15]. At this stage, however, the effect of the crystal structure or the atomic species on $f$ remains unclear. The systematic determination of $f$ for various semiconductors is expected to be effective for revealing the origin of the magnitude of $f$.

In this study, we have determined energy-dependent $f$ values of seven practical solar cell materials (GaAs, InP, CdTe, CuInSe$_2$ (CISe), CuGaSe$_2$ (CGSe), CZTSe and CZTS), in an attempt to characterize the band-edge light absorption properties. From our systematic DFT calculations, complete three-dimensional $f$ distributions in zincblende, chalcopyrite and kesterite Brillouin zones are visualized and compared to reveal the effects of the constituent atom and crystal structure. Quite surprisingly, we find that drastic increase of $f$ occurs in the $E_g$ region in all the compounds investigated here, which leads to the fact that $\alpha$ in the band-edge region is increased significantly by the giant near-edge $f$. Our result further indicates that atomic species does not affect $f$ and the tetrahedral atomic configuration is a key to interpret anomalous $f$ enhancement in the band edge region.

## II. DFT ANALYSIS

The DFT calculations were performed by generalized gradient approximation within the Perdew-Burke-Ernzerhof scheme (PBE) [16] using Advance/PHASE package. In PBE calculations, the structural optimization was made for all absorber crystals until the atomic configuration converged to within 5 meV/Å using a plane-wave cutoff energy of 455 eV. For the DFT calculation of zincblende crystals (i.e., GaAs, InP and CdTe), a two-atom primitive cell was used with a 10 × 10 × 10 $k$ mesh, while eight-atom primitive cells were employed for the calculations of chalcopyrite and kesterite semiconductors (CISe, CGSe, CZTSe and CZTS) with a 8 × 8 × 8 $k$ mesh.

The $\varepsilon_2$ spectrum of dielectric function ($\varepsilon = \varepsilon_1 - i\varepsilon_2$) is calculated according to

$$\varepsilon_2 = \sum_i \sum_j \varepsilon_2(V_i C_j), \tag{1}$$

where $\varepsilon_2(V_i C_j)$ shows the $\varepsilon_2$ contribution generated by the optical transition of $V_i \to C_j$. Here, $V_i$ and $C_j$ denote the $i$th valence band from the valence band maximum (VBM) and the $j$th conduction band from the conduction band minimum (CBM) in the $k$ space, respectively. The above equation shows that $\varepsilon_2$ is derived as a summation of $\varepsilon_2$ components obtained for each $V_i \to C_j$ transition. The $\varepsilon_2(V_i C_j)$ in Eq. (1) can be estimated explicitly by

$$\varepsilon_2(V_i C_j) = \frac{\hbar e^2}{8\pi^2 \varepsilon_0 m\omega} \int f_{ij} \delta(E_{C_j,\mathbf{k}} - E_{V_i,\mathbf{k}} - \hbar\omega) d\mathbf{k}, \tag{2}$$

where $E_{C_j}$ and $E_{V_i}$ show the energies of C$_j$ and V$_i$ in the $k$ space, respectively. The $m$, $\omega$, $V$ denote electron mass, angular frequency of incident light and volume, respectively. The $f_{ij}$ in the above equation represents the oscillator strength for the optical transition of V$_i \to$ C$_j$, defined by

$$f_{ij} = \frac{2m\omega}{\hbar}\left|\left\langle \Psi_{C_j}\left|\mathbf{u}\cdot\mathbf{r}\right|\Psi_{V_i}\right\rangle\right|^2, \qquad (3)$$

where $\left|\Psi_{C_j}\right\rangle$ and $\left|\Psi_{V_i}\right\rangle$ are the conduction and valence states, whereas **u** and **r** show the polarization vector and position operator, respectively. Eq. (3) indicates that the valence wave function is modified by the light excitation (i.e., polarization vector) and $f$ is derived from the overlap integration between the modified valence state (i.e., $\left|\mathbf{u}\cdot\mathbf{r}\right|\Psi_{V_i}\rangle$) and conduction state ($\left|\Psi_{C_j}\right\rangle$). As confirmed from Eq. (2), the $\varepsilon_2$ component is further determined by the integration of $f_{ij}$ at each transition energy (i.e., $\hbar\omega = E_{C_j} - E_{V_i}$). It should be noted that, since $\varepsilon_2 = 2nk$, where $n$ and $k$ are the refractive index and extinction coefficient, and $\alpha = 4\pi k/\lambda$, $\varepsilon_2$ increases as $\alpha$ becomes larger.

In actual dielectric function calculations, the spectra are calculated based on a method developed by Kageshima and Shiraishi [17]. For the calculations, we used a dense 12 × 12 × 12 $k$ mesh to suppress the distortion of calculated spectra [9]. In addition, for Cu-containing compounds, the onsite Coulomb interaction was considered for the Cu 3$d$ state [18] with an effective energy of $U_{\text{eff}} = 6$ eV.

In general, the DFT approximation within PBE underestimates $E_g$ seriously [19] and the energy position of the PBE-calculated $\varepsilon_2$ spectrum is red-shifted compared with the experimental spectrum. Thus, we have corrected the underestimated $\varepsilon_2$-spectral energy position by applying the sum rule [20]:

$$\int \varepsilon_2(E)E dE = const. \qquad (4)$$

Specifically, the sum rule requires that the amplitude of the $\varepsilon_2$ spectrum reduces to $E/(E+\Delta E)$ when the $\varepsilon_2$ spectrum is shifted toward higher energy by $\Delta E$ [21]. Accordingly, by using the parameter $\Delta E$, the underestimated contribution in the PBE approximation was corrected, so that the calculated DFT spectra match with those observed experimentally. The actual $\Delta E$ values employed in this study are 0.55 eV (GaAs), 0.50 eV (InP), 0.55 eV (CdTe), 0.36 eV (CISe), 0.33 eV (CGSe), 0.41 eV (CZTSe) and 0.40 eV (CZTS). The PBE-derived $E_g$ values have also been corrected by using the above $\Delta E$ values.

## III. RESULTS AND DISCUSSION

Figure 1 shows the optimized structures of (a) zincblende (GaAs), (b) chalcopyrite (CISe), (c) kesterite (CZTSe) crystals and the band structures of (d) GaAs, (e) CISe and (f) CZTSe, obtained from the DFT calculations. Although the atomic configurations of these crystals are different, all the structures have similar tetrahedral-bonding structures derived from the diamond crystal.

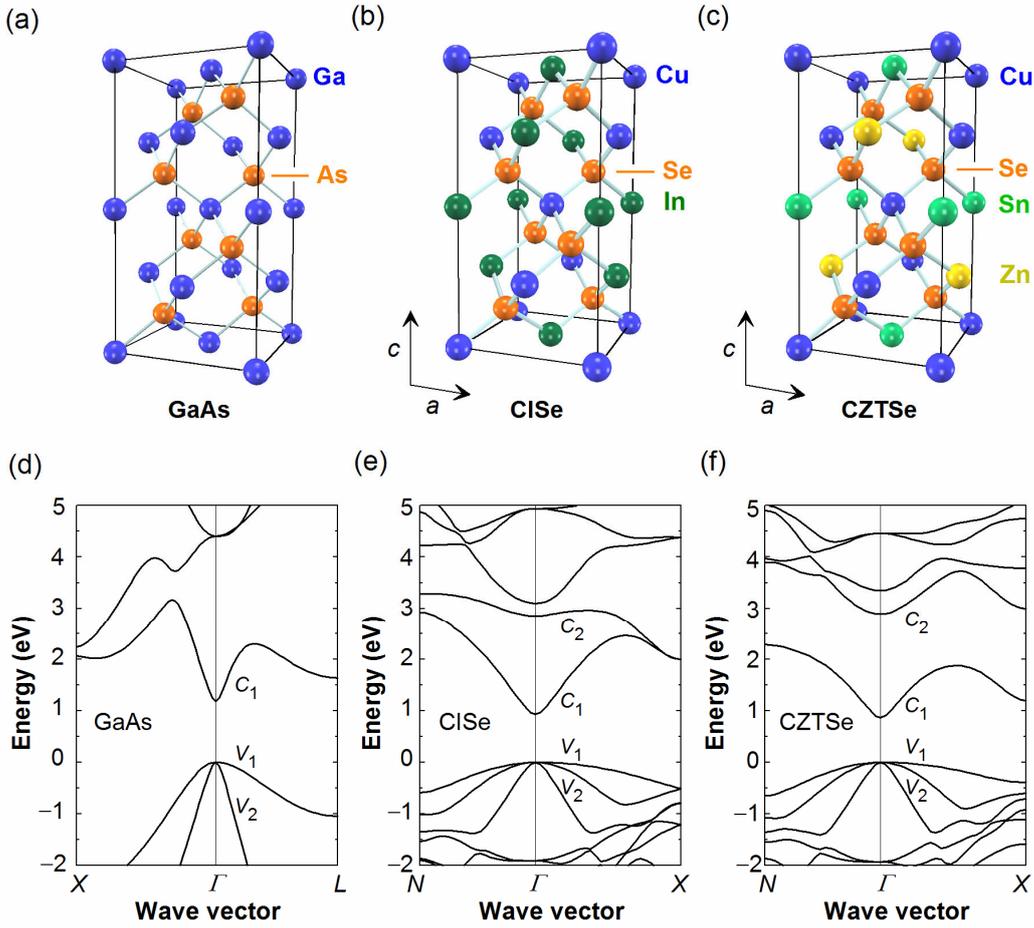

FIG. 1. Optimized structures of (a) GaAs (zincblende), (b) CISe (chalcopyrite), (c) CZTSe (kesterite) crystals and band structures of (d) GaAs, (e) CISe and (f) CZTSe, obtained from the DFT calculations. In (b) and (c), $a$ and $c$ denote the $a$ and $c$ axis directions. In (d)-(f), $V_j$ ($C_j$) represents the $j$th valence (conduction) band from valence band maximum (conduction band minimum).

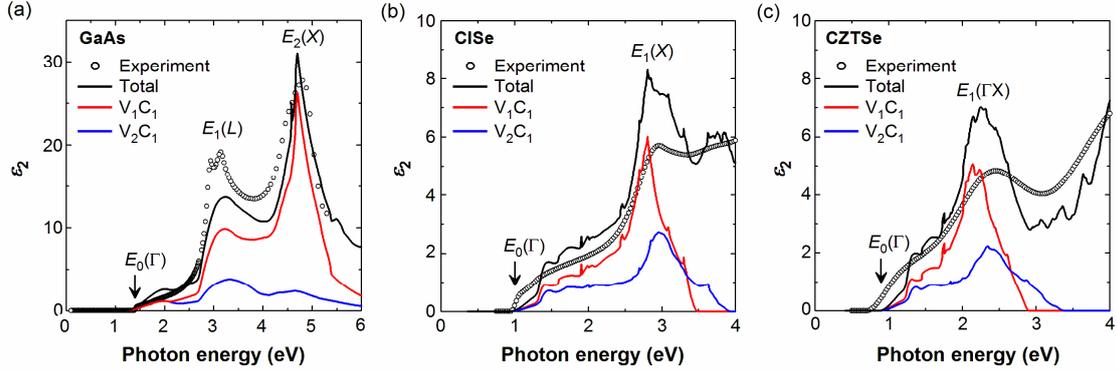

FIG. 2. $\varepsilon_2$ spectra of (a) GaAs, (b) CISe and (c) CZTSe. The experimental spectra (open circles) and DFT spectra (solid lines) are shown. The experimental spectra were taken from Ref. [2]. The DFT spectra have been blue-shifted with $\Delta E$. The $\varepsilon_2$ spectra indicated as "total" show the DFT spectra calculated including all the transitions, while $V_iC_j$ represents the $\varepsilon_2$ contribution for the optical transition from the $i$th valence band to the $j$th conduction band (i.e., $V_i \rightarrow C_j$).

In the band structures of Figs. 1(d)-(f), the energy positions of all the conduction bands are moved upward using the $\Delta E$ values mentioned above to correct the underestimated $E_g$ values (scissor operation) [22]. The band structures of the zincblende crystals (GaAs, InP and CdTe) are essentially similar [23], whereas the band structure of CISe is similar to those of CZTSe and CZTS [24].

Figure 2 shows the $\varepsilon_2$ spectra of (a) GaAs, (b) CISe and (c) CZTSe. The open circles indicate experimental spectra [2], whereas the solid lines show DFT spectra. All the calculated spectra have been blue-shifted according to the sum rule mentioned above. For the DFT spectra, the black lines indicate the total $\varepsilon_2$ spectra obtained from the summation of Eq. (1), while the red and blue lines indicate the $\varepsilon_2$ contributions for selected optical transitions. In particular, $V_iC_j$ in Fig. 2 represents the $\varepsilon_2$ component for the optical transition of $V_i \rightarrow C_j$, indicated in the band structures of Fig. 1.

In Fig. 2, the onset of the $\varepsilon_2$ spectra corresponds to the $E_0$ transition that originates from the optical transition at the $\Gamma$ point. In the case of GaAs, the two sharp peaks at 3.2 and 4.7 eV [i.e., $E_1$ and $E_2$ transitions in Fig. 2(a)] indicate the direct transitions at the L and X points in Fig. 1(d), respectively [25]. For CISe, the $E_1$ transition at 2.8 eV occurs at the X point in Fig. 1(e) [26], whereas the $E_1$ transition of CZTSe (2.2 eV) has been assigned to the transition at the midpoint between the $\Gamma$ and X points in Fig. 1(f) [27].

The $\varepsilon_2$ spectra of InP and CdTe are essentially similar to that of GaAs, while we also confirmed the similarity between the chalcopyrite crystals (CISe and CGSe) and the kesterite compounds (CZTSe and CZTS), as summarized in Supplementary Fig. 1.

Although the total $\varepsilon_2$ spectra are composed of numerous transitions that occur between $V_i$ and $C_j$, it can be confirmed from Fig. 2 that the low-energy $\varepsilon_2$ spectrum is determined primarily by the $V_1C_1$ transition [27]. In other words, the band-edge light absorption can be characterized by absolute $f$ values of the $V_1 \to C_1$ light absorption [i.e., $f_{11}$ in Eq. (2)]. In Fig. 2, $V_2C_1$ transition also contributes to the band-edge $\varepsilon_2$. This is based on the simple fact that $V_2$ overlaps with $V_1$ particularly near the $\Gamma$ point (see Fig. 1). In this study, we have mainly characterized $f_{11}$ to simply our DFT analysis and discussion.

Figure 3 shows (a) normalized three-dimensional $f_{11}$ distribution and (b) constant-energy $k$ surface (2.0-5.0 eV) for the $V_1C_1$ transition in the GaAs Brillouin zone. The numerical values in this figure indicate the energy separation between $E_C$ and $E_V$ in the $V_1 \rightarrow C_1$ transition (i.e., $E_{11} = E_{C_1} - E_{V_1}$) and thus $E_{11}$ simply corresponds to the energy difference between $C_1$ and $V_1$ bands in Fig. 1(d). For high symmetry planes in the Brillouin zone, the cross-sectional $f_{11}$ mapping images are also indicated. It can be seen that $f_{11}$ shows a unique three-dimensional distribution with $f_{11}$ at the Γ point being the highest. In other words, the transition probability increases drastically near the Γ point due to the large overlap between $\Psi_{C_1}$ and $\Psi_{V_1}$ in the optical excitation. As confirmed from Fig. 3(a), $f_{11}$ also shows high values between the Γ and L points in the Brillouin zone. Thus, in the range between the $E_0(\Gamma)$ to $E_1(L)$ transitions, the light absorption is characterized by high $f_{11}$.

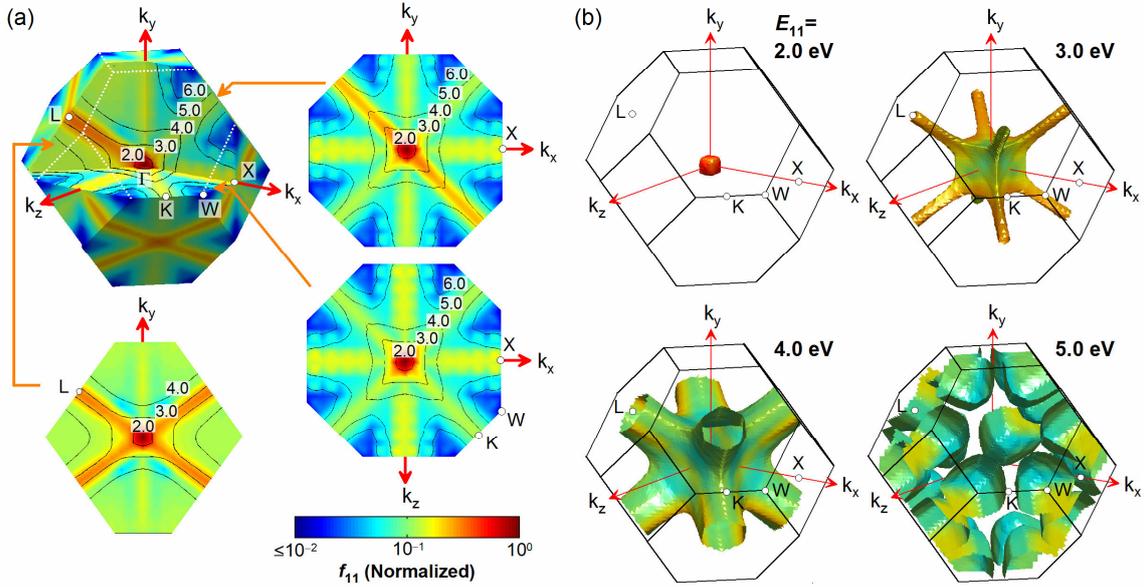

FIG. 3. (a) Normalized $f_{11}$ distribution and (b) constant-energy $k$ surface ($E_{11}$ = 2.0-5.0 eV) for the $V_1C_1$ transition in the GaAs Brillouin zone. In (a), the numerical values indicate the energy separation between $E_C$ and $E_V$ in the $V_1 \to C_1$ transition (i.e., $E_{11} = E_{C_1} - E_{V_1}$). In (b), the $f_{11}$ distributions on different $E_{11}$ energy surfaces are indicated.

In Fig. 3(b), the constant-energy $k$ surfaces, corresponding to $E_{11}$ of 2.0, 3.0, 4.0 and 5.0 eV, are shown. The $k$-surface shape is cubic near the Γ point but the $k$ surface at high $E$ extends toward the L point, showing complicated structures. In Fig. 3(b), the corresponding $f_{11}$ distribution indicated in Fig. 3(a) is also overlapped on the constant-energy $k$ surface. As shown in Fig. 3(b), the transition energy (i.e., $E_{11}$) determines the shape of the $k$ surface and, if all the $f_{11}$ values on this particular constant-$k$ surface are integrated, the $\varepsilon_2$ component for the selected transition energy is estimated [see Eq. (2)]. Thus, $\varepsilon_2$ increases when the $f_{11}$ values on the constant-energy $k$ surface are larger and the area of the $k$ surface is wider. Near the $E_g$ region, the area of the $k$ surface is small but the high $f_{11}$ increases $\varepsilon_2$ in this range. When the energy is high, the area of the constant-$k$ surface becomes quite large but with the lower $f_{11}$ values. At high energies, therefore, $\varepsilon_2$ is more governed by the total area of the $k$ surface. It should be noted that the analysis of Fig. 3 corresponds to the result for the $V_1C_1$ transition and, if $\varepsilon_2$ values for all the $V_iC_j$ transitions are summated, the absolute $\varepsilon_2$ value is determined, as shown in Eq. (1). However, the $f$ distribution of all the summated transitions is still similar to that of Fig. 3(a) (see Supplementary Fig. 2).

Figure 4 shows the normalized $f_{11}$ distributions of (a) GaAs, (b) InP, (c) CdTe, (d) CISe and (e) CZTSe Brillouin zones. In Fig. 4(a)-(c), the Brillouin zone obtained from a standard zincblende structure is shown, while the shape of the CISe and CZTSe Brillouin zones represent that obtained from an eight-atom primitive cell and is derived from the zone folding of the zincblende Brillouin zone [1]. In Fig. 4(f), for comparison, the $f_{11}$ distribution of CdTe calculated assuming an eight-atom primitive cell is also shown.

In Fig. 4, the overall $f_{11}$ values of GaAs, InP and CdTe show quite similar trends,

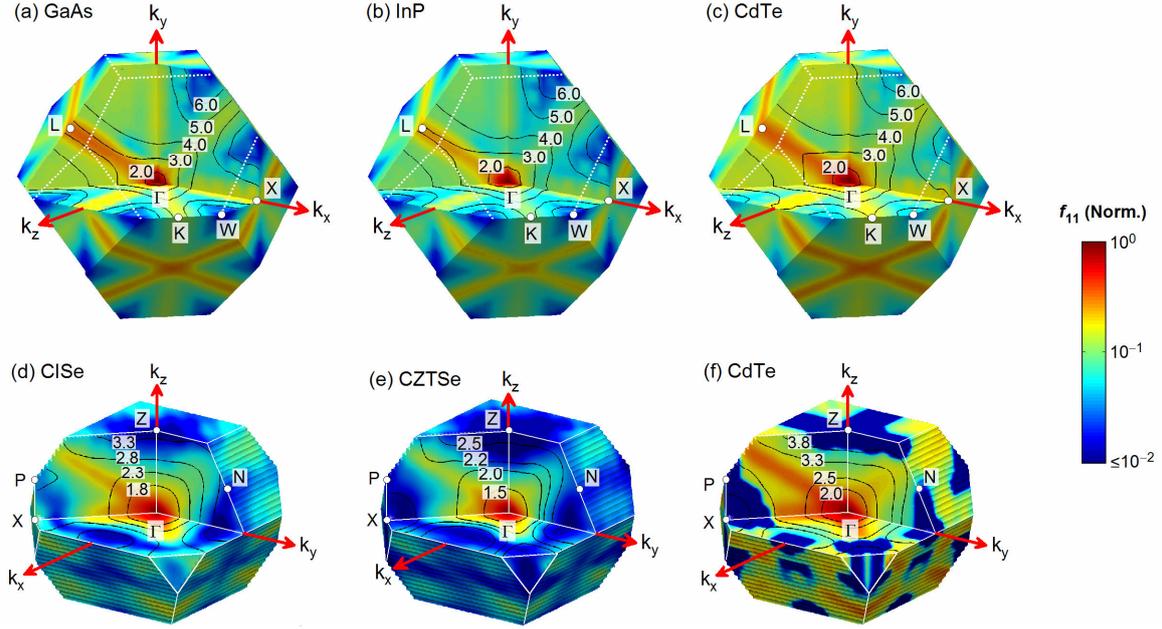

FIG. 4. Normalized $f_{11}$ distributions of (a) GaAs, (b) InP, (c) CdTe, (d) CISe and (e) CZTSe Brillouin zones. The Brillouin zones obtained from standard zincblende structures [(a)-(c)] and eight-atom primitive cell structures [(d)-(f)] are shown. In (f), for comparison, the $f_{11}$ distribution of CdTe calculated assuming an eight-atom primitive cell is shown.

indicating clearly that the $f_{11}$ distribution is independent of atomic species and is determined by the crystal structure. The comparison of $f_{11}$ in Fig. 4(d)-(f) further confirms that the $f_{11}$ distributions of zincblende, chalcopyrite and kesterite crystals are essentially the same and all the structures show quite high $f_{11}$ near the Γ point. In other words, near band-edge optical transitions in all the crystals are modified significantly by the anomalous increase in $f_{11}$ at the Γ point. We further confirmed that the $f_{11}$ distributions of CGSe and CZTS are quite similar to those of CISe and CZTSe in Fig. 4 (Supplementary Fig. 3).

To discuss the significant variation in $f_{11}$ observed in the Brillouin zones, we further calculated absolute $f_{11}$ and joint density-of-states ($J_{CV}$) for the $V_1C_1$ transition. As known well [20], $J_{CV}$ is given by

$$J_{CV}(E_{T,\mathbf{k}}) = \frac{1}{4\pi^3} \int \frac{dS_{\mathbf{k}}}{|\nabla_{\mathbf{k}}(E_{T,\mathbf{k}})|}, \qquad (5)$$

where $E_{T,\mathbf{k}}$ shows the transition energy (i.e., $E_{T,\mathbf{k}} = E_{C,\mathbf{k}} - E_{V,\mathbf{k}}$) and $S_{\mathbf{k}}$ is the constant energy surface defined by $E_{T,\mathbf{k}}$ = const. As confirmed from the above equation, $J_{CV}$ essentially shows the volume change of the constant-energy $k$ surface and $J_{CV}$ increases when the volume change per energy becomes large. Here, we define $J_{CV}$ corresponding to the $V_1C_1$ transition as $J_{11}$.

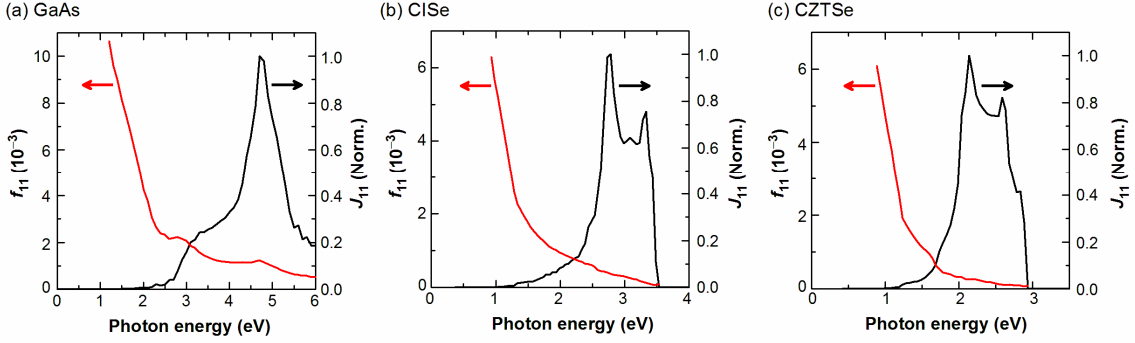

FIG. 5. $f_{11}$ and $J_{11}$ spectra of (a) GaAs (zincblende), (b) CISe (chalcopyrite) and (c) CZTSe (kesterite) crystals. For $J_{11}$, normalized values are shown.

Figure 5 summarizes $f_{11}$ and $J_{11}$ spectra of (a) zincblende (GaAs), (b) chalcopyrite (CISe) and (c) kesterite (CZTSe) crystals. It can be seen that $f_{11}$ of all the semiconductors in Fig. 5 shows steep increase in the band-edge region. Such giant oscillator strength in the band-edge optical transition has been overlooked completely in earlier studies.

In the case of GaAs, $f_{11}$ in the band edge is more than one order of magnitude larger than that in a high $E$ region ($E > 4$ eV). In particular, in the near-$E_g$ region ($E < 2$ eV), $J_{11}$ is very small and the light absorption is primarily governed by the large magnitude of $f_{11}$. In the high $E$ region ($E > 2$ eV), the $J_{11}$ spectrum of GaAs resembles to the corresponding $\varepsilon_2$ spectrum [see Fig. 2(a)]. Specifically, in the GaAs $J_{11}$ spectrum, the shoulder at 3.2 eV and the transition peak at 4.7 eV correspond to the $E_1$ and $E_2$ transitions, respectively. Thus, the high-energy $\varepsilon_2$ spectrum is determined predominantly by $J_{11}$, as the volume change of $S_{\mathbf{k}}$ for $E_{T,\mathbf{k}}$ is large in this region [see Fig. 3(b)]. In an earlier study, a $J_{CV}$ spectrum similar to Fig. 5(a) has also been reported [7].

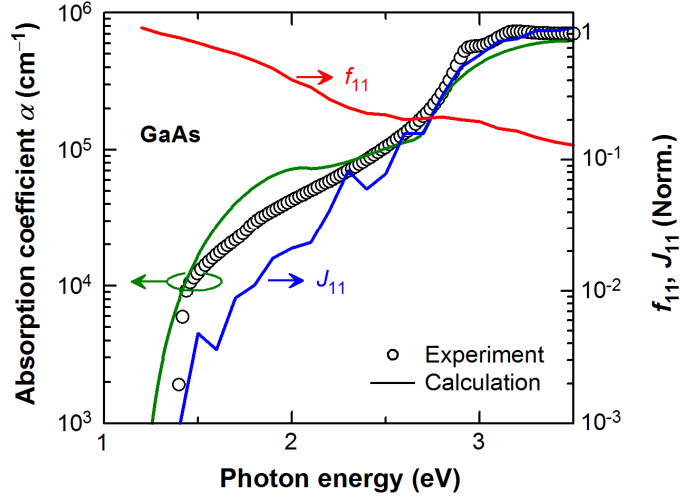

FIG. 6. $\alpha$ spectra of GaAs obtained from the experiment (open circles) and the DFT calculation (solid line), together with the normalized $f_{11}$ and $J_{11}$ in logarithmic scales. The experimental spectrum was taken from Ref. [2].

The spectral features of CISe and CZTSe in Figs. 5(b) and (c) are quite similar even though the energy positions are different. Similar to GaAs, $f_{11}$ of CISe and CZTSe is quite high in the $E_g$ region, while very low $J_{11}$ is confirmed in the same region. The sharp peaks observed in $J_{11}$ at 2.8 eV (CISe) and 2.2 eV (CZTSe) are consistent with the $E_1$ transition peaks in their $\varepsilon_2$ spectra. For the distributions of $f_{11}$ and $J_{11}$, we confirmed the similarities among the zincblende crystals (GaAs, InP and CdTe) and between the chalcopyrite compounds (CISe and CGSe) and kesterite compounds (CZTSe and CZTS), as shown in Supplementary Fig. 4.

All the above results indicate the presence of the universal features for the light absorption phenomenon in practical solar-cell absorbers, which can be categorized into two regions: namely, (i) $f_{11}$-dominated optical transition region at low energies, which determines the band-edge $\alpha$ of the absorber, and (ii) $J_{11}$-dominated transition region at high energy, which characterizes the overall shape of the $\varepsilon_2$ spectra. Accordingly, general understanding that $J_{CV}$ determines the overall light absorption is not accurate and the effect of $f$ on the optical transition is significant near $E_g$. To elucidate this effect, in Fig. 6, the $\alpha$ spectra of GaAs are shown, together with the normalized $f_{11}$ and $J_{11}$ in logarithmic scales. In this figure, the DFT $\alpha$ spectrum (solid line) shows excellent agreement with the experimental spectrum (open circles), taken from Ref. [2]. In the band-edge region, $J_{11}$ reduces drastically, while $f_{11}$ indicates a large increase. Thus, the result of Fig. 6 confirms the important role of $f_{11}$ in determining the band-edge $\alpha$ values.

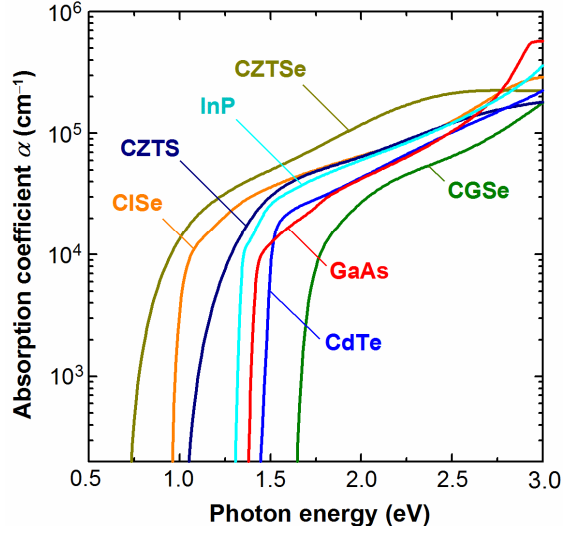

FIG. 7. Experimental $\alpha$ spectra of GaAs, InP, CdTe, CISe, CGSe, CZTSe and CZTS. The data were taken from Refs. [1,2]. All the materials show similar $\alpha$ values of $\sim 10^4$ cm$^{-1}$ in the energy region slightly above $E_g$.

Figure 7 shows the experimental $\alpha$ spectra of the semiconductors [1,2], investigated in this study. It is evident that all the materials show similar $\alpha$ values of $\sim 10^4$ cm$^{-1}$ in the energy region slightly above $E_g$. Similar $\alpha$ values observed for all the tetrahedrally bonded semiconductors are consistent with our analyses. Specifically, similar $f_{11}$ magnitudes have been confirmed for all the absorbers in Fig. 7.

The above analysis results reveal a quite important fact that $f$ is governed primarily by the fundamental crystal structure and not by atomic species. This conclusion is surprising as the origin of $\Psi_C$ and $\Psi_V$ changes drastically with atomic species. In the case of GaAs, for example, the VBM and CBM regions consist mainly of As(4$p$) and Ga(4$s$,4$p$)+As(4$p$), respectively, with As anions of the VBM being in bonding states. In contrast, the VBM of CISe and CZTSe is characterized by the anti-bonding states formed by the interaction of Cu(3$d$) and Se(4$p$) [28], while the CBM mainly consists of cation $s$ and anion $sp$ orbitals [29]. Quite interestingly, all these details concerning $\Psi_C$ and $\Psi_V$ do not affect $f_{11}$ and the absolute $\alpha$ near $E_g$ is independent of atomic species as long as the fundamental crystal structure (tetrahedral atomic configuration) is maintained. This can be the reason why many practical solar cells are based on tetrahedrally bonded crystals.

## IV. CONCLUSION

The oscillator strength distributions of seven practical solar-cell materials (GaAs, InP, CdTe, CISe, CGSe, CZTSe and CZTS) have been calculated based on DFT to reveal the nature of the optical transition in these semiconductors. In particular, to understand the light absorption characteristic near the band-edge region, $f$ and $J_{CV}$ for the transition from the first valence band to the first conduction band are evaluated. We found that $f_{11}$ of all the crystals investigated here increases drastically near the $\Gamma$ point in the Brillouin zone and the anomalous increase of $f_{11}$ enhances $\alpha$ significantly in a low energy region near $E_g$. The giant $f_{11}$ observed for zincblende, chalcopyrite and kesterite structures originates from the unique tetrahedral-bonding structure with negligible contribution of atomic species. In a high energy region above $E_g$, $f_{11}$ reduces steeply while $J_{11}$ increases. Based on the DFT calculations, we established that the optical transitions in solar-cell light absorbers can be categorized into two regions with (i) the $f_{11}$-dominated absorption region near the band-edge and (ii) the $J_{11}$-dominated absorption region in a high energy region. In contrast to general understanding that the band-edge light absorption is derived from $J_{CV}$, our result reveals the critical role of $f_{11}$ in the determination of absolute $\alpha$ values near $E_g$ in quite general semiconductor materials.

**Supplementary Information**

**Giant oscillator strength in the band-edge light absorption of zincblende, chalcopyrite and kesterite solar cell materials**

Masato Kato, Mitsutoshi Nishiwaki and Hiroyuki Fujiwara[*]

*Department of Electrical, Electronic and Computer Engineering, Gifu University, 1-1 Yanagido, Gifu 501-1193, Japan*

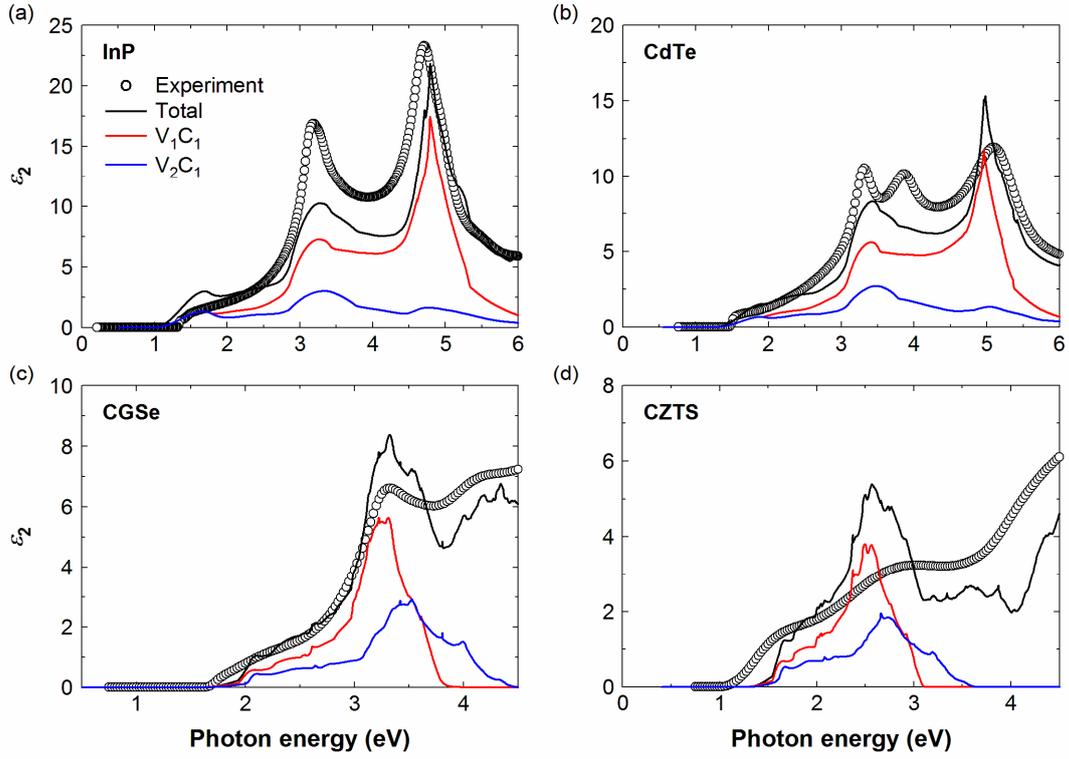

Supplementary FIG. 1. $\varepsilon_2$ spectra of (a) InP, (b) CdTe, (c) CGSe, and (d) CZTS. The experimental spectra (open circles) and DFT spectra (solid lines) are shown. The DFT spectra have been blue-shifted with $\Delta E$. The $\varepsilon_2$ spectra indicated as "total" show the DFT spectra calculated including all the transitions, while $V_iC_j$ represents the $\varepsilon_2$ contribution for the optical transition from the $i$th valence band to the $j$th conduction band (i.e., $V_i \rightarrow C_j$).

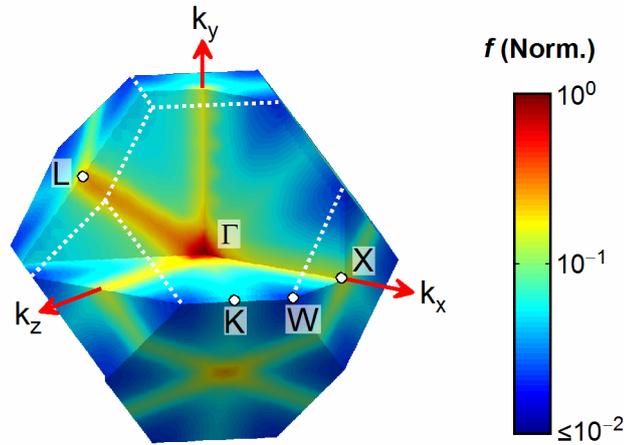

Supplementary FIG. 2. Normalized $f$ distribution for all the optical transitions in the GaAs Brillouin zone.

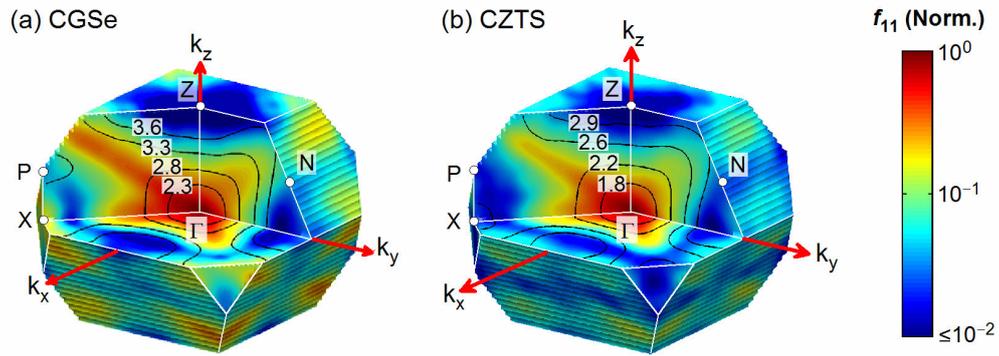

Supplementary FIG. 3. Normalized $f_{11}$ distribution in (a) CGSe and (b) CZTS Brillouin zones. The Brillouin zones obtained from eight-atom primitive cell structures are shown.

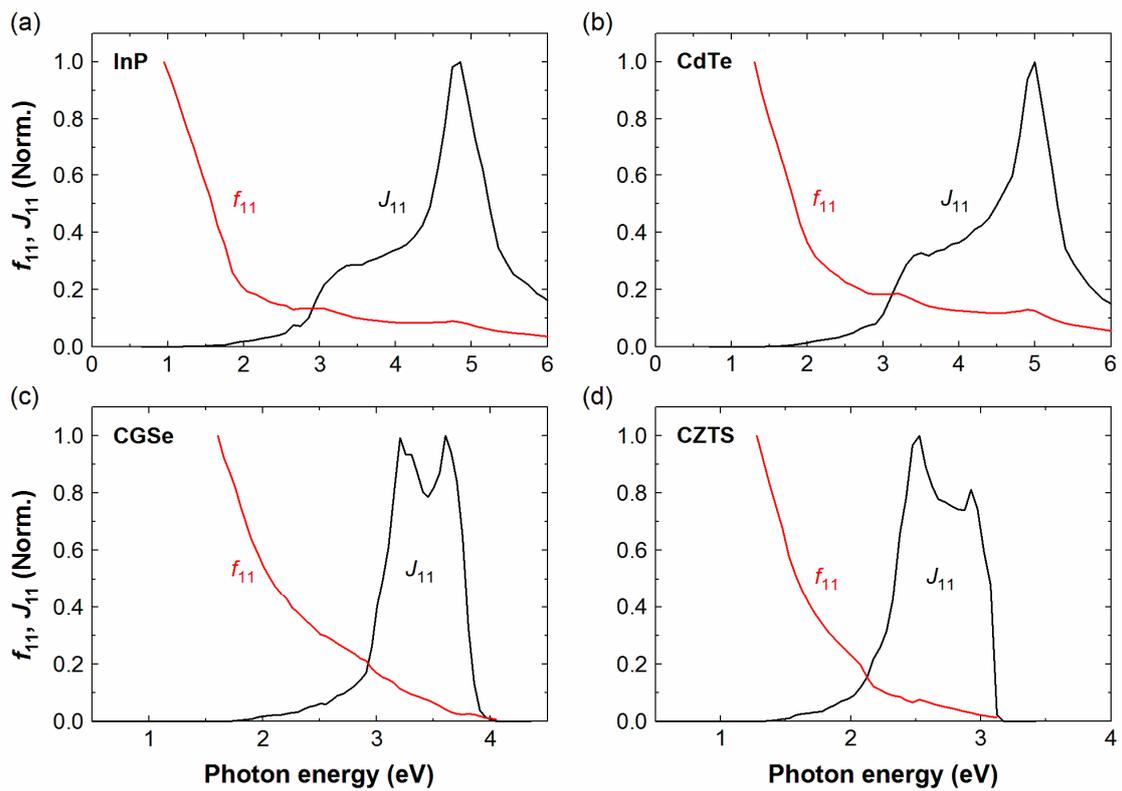

Supplementary FIG. 4. $f_{11}$ and $J_{11}$ spectra of (a) InP, (b) CdTe, (c) CGSe and (d) CZTS crystals. The normalized values are shown.